\documentclass[%
reprint,
superscriptaddress,
 amsmath,amssymb,
 aps,
pra,
]{revtex4-2}
\usepackage{amsmath,amsfonts,amssymb}
\usepackage{graphicx}
\usepackage{dcolumn}
\usepackage{bm}

\usepackage[dvipsnames]{xcolor}  
\usepackage{hyperref}              
\hypersetup{
	colorlinks=true,
	allcolors=black                
}
\usepackage{orcidlink}
\usepackage{ulem}
\usepackage{siunitx}

\def\e{\text{e}}
\def\ii{\text{i} }

\newcommand{\AiTr}{{{\text{Ai}_\text{Tr}}}}
\newcommand{\FTAiTr}{{\widetilde{\text{Ai}}_\text{Tr}}}

\newcommand{\Ai}{{{\text{Ai}}}}

\begin{document}

\preprint{APS/123-QED}

\title{Correlated Quantum Airy Photons: An Analytical Approach}

\author{V. Sau}
\affiliation{Physics, Indian Institute of Technology Patna, Patna, 801106, Bihar, India}

\author{R. Giustozzi}
\affiliation{Physics Division, School of Science and Technology, University of Camerino, I-62032 Camerino, MC, Italy}

\author{P. Piergentili \orcidlink{0000-0003-2125-4766}}
\affiliation{Physics Division, School of Science and Technology, University of Camerino, I-62032 Camerino, MC, Italy}
\affiliation{INFN, Sezione di Perugia, via A. Pascoli, I-06123 Perugia, Italy}

\author{D. Vitali \orcidlink{0000-0002-1409-7136}}
\affiliation{Physics Division, School of Science and Technology, University of Camerino, I-62032 Camerino, MC, Italy}
\affiliation{INFN, Sezione di Perugia, via A. Pascoli, I-06123 Perugia, Italy}
\affiliation{CNR-INO, Largo Enrico Fermi 6, I-50125 Firenze, Italy}

\author{G. Di Giuseppe \orcidlink{0000-0002-6880-3139}}
\affiliation{Physics Division, School of Science and Technology, University of Camerino, I-62032 Camerino, MC, Italy}
\affiliation{INFN, Sezione di Perugia, via A. Pascoli, I-06123 Perugia, Italy}

\author{S. Ghosh}
\affiliation{Physics, Indian Institute of Science Education and Research Kolkata, Kolkata, 741246, West Bengal, India}

\author{U. Roy \orcidlink{0000-0003-2500-078X}}
\affiliation{Physics, Indian Institute of Technology Patna, Patna, 801106, Bihar, India}

\date{\today}

\begin{abstract}
We describe the generation of correlated photon pairs by means of spontaneous parametric down-conversion of an optical pump in the form of a finite energy Airy beam. The optical system function, which contributes to the propagation of the down-converted beam before being registered by the detectors, is computed. The spectral function is utilized to calculate the biphoton amplitude for finding the coincidence count of the inbound Airy photons in both far-field and near-field configurations. We report the reconstruction of the finite energy Airy beam in the spatial correlation of the down-converted beams in near field scenario. In far field, the coincidence counts resembles the probability density of the biphoton in momentum space, revealing a direct mapping of the anti-correlation of the biphoton momentum. By examining the spatial Schmidt modes, we also demonstrate that longer crystals have tighter real-space correlations, but higher-dimensional angular correlations, whereas shorter crystals have fewer modes in momentum space and broader multimode correlations in position space.
\end{abstract}

\maketitle

\section{\label{sec:level1}Introduction}

The use of non--diffracting or quasi non-diffracting beams~\cite{durnin1987exact,gutierrez2000alternative} has attracted an increasing interest in optical and atomic physics due to their potential usefulness for circumventing the loss due to a finite distance propagation. Berry and Balazs showed that there exists a non-trivial solution 
of the free particle Schr\"{o}dinger equation which exhibits remarkable properties like self-acceleration and non--diffraction~\cite{berry1979nonspreading,unnikrishnan1996uniqueness}. In spite of its intriguing properties, the Airy wave packet was considered unrealistic due to its infinite energy. However, a truncated Airy beam was experimentally investigated in 2007~\cite{Siviloglou:07,siviloglou2007observation}. Interestingly, this finite energy Airy beam also exhibits non-diffracting and self-accelerating properties up to a considerable distance. This has attracted enormous attention both theoretically and experimentally. In fact, a number of emerging areas such as communication \cite{zhang2014modulation,rose2013airy}, super-resolution imaging \cite{jia2014isotropic}, micro-particle manipulation \cite{singh2017particle}, parabolic plasma channeling \cite{polynkin2009curved}, bio-medical application \cite{vettenburg2014light}, might benefit from it.
To model the propagation of a beam in a non-linear medium, which is non--diffracting in nature, is quite non-trivial \cite{roy2011propagation}. It has not been investigated in details whether the non--diffracting nature of the Airy beam when passed through a non-linear crystal, can aid in quantum advantages by giving birth to a non-diffracting biphoton \cite{mclaren2014self}.

Owing to the exotic properties, an entangled Airy beam could be useful in quantum information schemes including distortion--free quantum communication. For this purpose and various quantum applications, it is interesting to study how these entangled photons can be generated and manipulated. Spontaneous parametric down--conversion (SPDC) is the most widely used scheme to produce these entangled photon pairs and also single photon sources \cite{couteau2018spontaneous,de2006non,atature2002multiparameter}. The down converted beams can be entangled in different degrees of freedom, such as frequency, momentum, position and polarization~\cite{di2002entangled}, and they are routinely used in quantum information and technology applications \cite{jennewein2000quantum,angeletti2023microwave}. This can also serve as a tool for engineering various quantum states such as Schr\"{o}dinger cat-like state \cite{de1999generating}, hyper-entangled state~\cite{atature2002multiparameter}, and for their implementation as high dimensional entangled states to increase the security and information encoding capacity in a quantum key distribution task \cite{scarani2009security}. These mesoscopic states, like Schr\"{o}dinger cat state, compass state and others, hold the key for generating sub--Planck scale structures which takes the quantum precision measurements to newer heights \cite{agarwal2004mesoscopic,panigrahi2011sub,bera2022quantum,batin2024quantum}. Recently, using entangled optical beams, an improvement in the sensitivity of force measurement has also been reported \cite{di2023entangled,Xi2023}. In addition, there has been a plethora of studies in which the transverse spatial and momentum correlation of an entangled beam after parametric down-conversion has been explored both theoretically and experimentally \cite{jost1998spatial,strekalov1995observation,pittman1996two,rubin1996transverse,castelletto2005spatial,schneeloch2016introduction,fedorov2008spontaneous,walborn2007transverse,reichert2017quality,chan2007transverse,tasca2009propagation}. Using an Airy pump beam in SPDC, one can get a high correlation between two down-converted beams \cite{maruca2018quantum,aadhi2016airy}. In this work, we adopt an analytical approach to study the spatial transverse correlation of the Airy biphoton.  We first discuss the use of an experimentally generated Airy beam in the generation of correlated photon pairs by means of SPDC. Then the propagation of the beam inside the nonlinear crystal as well as in free space is modelled. We analyze the optical system functions through which the down-converted beam travels before being detected. By evaluating the spectral function of the two-photon state, we investigate the biphoton amplitude and coincidence-counting rate for the incoming photons in both far-field and near-field configurations. The description provided here shows that the entangled photon pairs generated by an Airy beam pump may represent a powerful tool for low-loss quantum communications applications. 

The paper is organized as follows. In Sec.~II, the basic elements of the analytical description are provided, while in Sec.~III, the field evolution to the detection stage is described. In Sec.~IV, the behavior of the generated entangled beams is described both in the far-field and in the near field, while in Sec.~V, the spatial entanglement capabilities of the generation scheme are investigated. Sec.~VI is for concluding remarks. 

\section{Analytical Model}

\begin{figure}
\includegraphics[width=.4\textwidth]{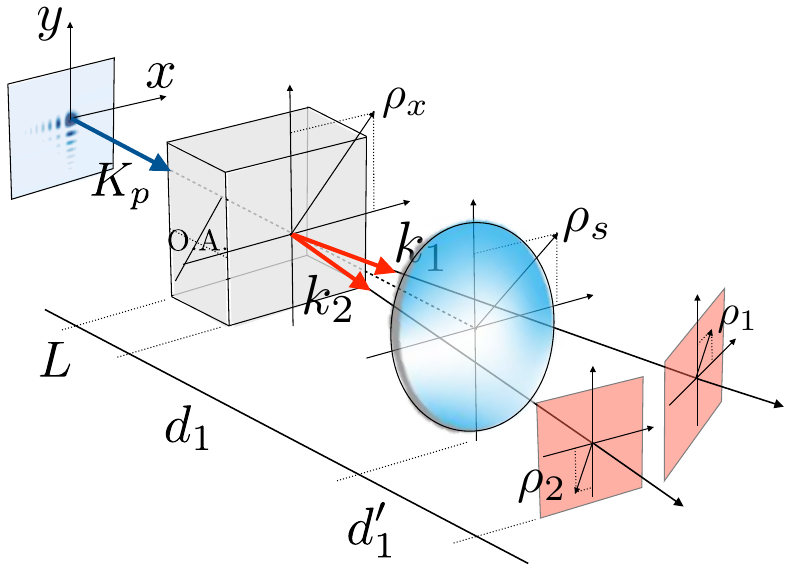}
\caption{Schematic diagram of the photon pairs generation. A pump beam at frequency $\omega_p$ and wave vector $\bm{K}_p$ travels through a non--linear crystal of length $L$, producing pairs of correlated photons at $(\omega_1, \bm{k}_1)$ and $(\omega_2, \bm{k}_2)$. A lens at plane $\bm{\rho}_s$ is placed at distance $d_1$ from the crystal output plane $\bm{\rho}_x$, and $d_2$ from the detection planes $\bm{\rho}_j$. O.A. indicates the crystal optical axes, which has component along $\hat z$ and $\hat y$, the extraordinary direction, while the $\hat x$ is the ordinary one.}
	\label{schematic}
\end{figure}

Spontaneous parametric down-conversion is a non-linear process in which an incoming photon of higher frequency splits into a pair of twin photons after passing through a non-centrosymmetric crystal due to its birefringence property.
We consider the system shown in Fig.~\ref{schematic}, where an input single--frequency pump beam travels through a non--linear crystal of thickness $L$. After emerging from the crystal, signal and idler photons go through a convex lens at the $ \bm{\rho}_{s}$ plane before being detected at $\bm{\rho}_{1}$ and $ \bm{\rho}_{2}$ plane.

The effective interaction Hamiltonian for the down-conversion process can be written as
\begin{equation}\label{hamiltonian}
H_{I}=\epsilon_{0}\int_{V}d^{3}r\chi^{(2)} E_{p}^{+}E_{1}^{-}E_{2}^{-}+ \text{H.C.},
\end{equation}
where H.C. stands for Hermitian conjugate, $V$ is the volume of the crystal covered by input beam, $\epsilon_{0}$ is the vacuum permittivity, $\chi^{(2)}$ is the second order nonlinear response of the medium, and $E_{p}^{+}$  is the electric field for positive frequency part of the input beam, which can be treated as classical, written as a superposition of plane waves having different weight factors, the angular spectrum of the beam, and given by

\begin{equation}
	E_{p}^{+}=\int d\bm{\kappa}_{p}\widetilde{E_p}(\bm{\kappa}_{p})
	\e^{\ii({k}_{pz}z+\bm{\kappa}_{p}\cdot \bm{\rho}-\omega_{p}t)}.
\end{equation}
Here, $\bm{\kappa}_{p}$ and $\bm{\rho}$ are the transverse momentum wave vector and transverse position vector, respectively. $\widetilde{E_p}(\bm{k}_{p})$ is the spatial amplitude of the pump beam in transverse momentum space, and ${k}_{pz}$ is the longitudinal wave number of the pump beam along the $z$-direction. $E_{1}^+$ and $E_{2}^+$ are the field operators:
\begin{equation}
	E_{j}^{+}=\sum_{\bm{k}}a_{j\bm{k}}\e^{\ii(\bm{k}\cdot\bm{r}-\omega_{\bm{k}j}t)}\,,
\end{equation}
where we omit a quantization factor, and $a_{j\bm{k}}$ is the annihilation operator of the $j-$polarized mode with wave vector $\bm{k}$~\cite{fox2006quantum}. For small non--linear interaction response, the radiation at the output face of the crystal can be evaluated by means of a perturbative approach, which, at first--order, provides

\begin{equation}\label{ip}
	|\psi \rangle=|0 \rangle+ \frac{1}{\ii\hbar} \int dt H_{I}|0\rangle,
\end{equation}
where $|0\rangle$ is the vacuum state, $H_{I}$ is the effective Hamiltonian in interaction picture. In the paper, we consider the limits of integration run from $-\infty$ to $+\infty$, unless it is specified otherwise. By using the expression of the Hamiltonian~\eqref{hamiltonian} in Eq.~\eqref{ip}, the state $|\psi \rangle$ can be written as \cite{rubin1994theory}

\begin{equation}\label{psi}
|\psi\rangle=|0\rangle+\sum_{\bm{k}_{1},\bm{k}_{2}}F(\bm{k}_{1},\bm{k}_{2})a^{\dag}_{\bm{k}_{1}}a^{\dag}_{\bm{k}_{2}}|0\rangle,
\end{equation}
where $a^{\dag}_{j\bm{k}}$ is the creation operator for mode $(j,\bm{k})$. The spectral function $F(\bm{k}_{1},\bm{k}_{2})$ can be cast as~\cite{pittman1996two,klyshko2018photons}
\begin{equation}
	F(\bm{k}_{1},\bm{k}_{2})=\langle 0|a_{\bm{k}_{1}}a_{\bm{k}_{2}}|\psi \rangle\,,
\end{equation}
and calculated by using Eq.~\eqref{ip}. We note that the time integral in Eq.~\eqref{ip} provides $\delta(\omega_{p}-\omega_{\bm{k}_{1}}-\omega_{\bm{k}_{2}})$, which indicates the energy conservation. The integral over the interaction volume can be separated into a longitudinal contribution over the crystal length $L$, and a transverse one over the illuminated area of the crystal. Finally we have
\begin{equation}\label{SF1}
	F(\bm{k}_{1},\bm{k}_{2})=\Gamma \delta (\omega_{p}-\omega_{\bm{k}_{1}}-\omega_{\bm{k}_{2}})h(L\Delta)
	H_{tr}(\bm{\kappa}_{1},\bm{\kappa}_{2}),
\end{equation}
where $\Gamma$, known as parametric gain index~\cite{rubin1996transverse}, includes all the constant factors.
The function $h$ is given by
\begin{equation}
	h(L\Delta)=\int_{-L}^{0}dz \e^{i\Delta z}=\frac{1-\e^{-\ii L\Delta}}{\ii L\Delta} \,,
\end{equation}
where, $\Delta={k}_{pz}-k_{1z}-k_{2z}$ represents the longitudinal wave number mismatch.
The transverse contribution $H_{tr}(\bm{\kappa}_{1},\bm{\kappa}_{2})$ with $\bm{\kappa}_j$, the transverse momentum wave vector, can be evaluated by assuming the area of the crystal illuminated by the pump beam much smaller than the cross sectional area of the crystal. Thus, limits in the surface integral effectively extends over an infinite range giving us a delta function, $\delta (\bm{\kappa}_{p}-\bm{\kappa}_{1}+\bm{\kappa}_{2})$, which shows the phase matching condition in transverse wave vector, \textit{i.e.} $\bm{\kappa}_{p}=\bm{\kappa}_{1}+\bm{\kappa}_{2}$. Finally the transverse integral takes the form
\begin{align}
	H_{tr}(\bm{\kappa}_{1},\bm{\kappa}_{2})
		=\widetilde{E_p}(\bm{\kappa}_{1}+\bm{\kappa}_{2})\,,
\end{align}
where, $\widetilde{E_p}(\bm{\kappa}_{1}+\bm{\kappa}_{2})$ is the Fourier transform of the pump beam profile, and the spectral function reads
\begin{align}\label{SF2}
	F(\bm{k}_{1},\bm{k}_{2})=\Gamma \delta (\omega_{p}-\omega_{\bm{k}_{1}}-\omega_{\bm{k}_{2}})
		\phi(\bm{k}_{1},\bm{k}_{2})\,.
\end{align}
where we introduce the function
\begin{align}\label{phi}
	\phi(\bm{k}_{1},\bm{k}_{2}) =
		\widetilde{E_p}(\bm{\kappa}_{1}+\bm{\kappa}_{2})
		\text{sinc}\left[\frac{L\Delta(\bm{k}_{1},\bm{k}_{2})}{2}\right]
		\e^{-\ii\frac{L\Delta(\bm{k}_{1},\bm{k}_{2})}{2}}\,,
\end{align}

\subsection{Type--I phase  matching}
The longitudinal phase mismatch term, $\Delta$ is a function of the frequencies $\omega_p$, $\omega_1$, and $\omega_2$, and transverse momenta $\bm{\kappa_p}$, $\bm{\kappa_1}$, and  $\bm{\kappa_2}$, and can be written as
\begin{align}
	\Delta = \sqrt{k_p(\omega_p,\bm{\kappa}_p) -|\bm{\kappa}_p|^2}
		&- \sqrt{k_1(\omega_1,\bm{\kappa}_1) - |\bm{\kappa}_1|^2} +
			\nonumber\\
		&-\sqrt{k_2(\omega_2,\bm{\kappa}_2) - |\bm{\kappa}_2|^2}
			\,.
\end{align}
with $k(\omega,\bm{\kappa}) = \omega\, n(\omega,\bm{\kappa})/c$, where $n(\omega,\bm{\kappa})$ is the index of refraction, and the constraints $\omega_p = \omega_1 + \omega_2$, and $\bm{\kappa}_p = \bm{\kappa}_1 + \bm{\kappa}_2$ are obeyed. It depends specifically on the non--linear birefringent medium.

\begin{figure}
\includegraphics[width=0.4\textwidth]{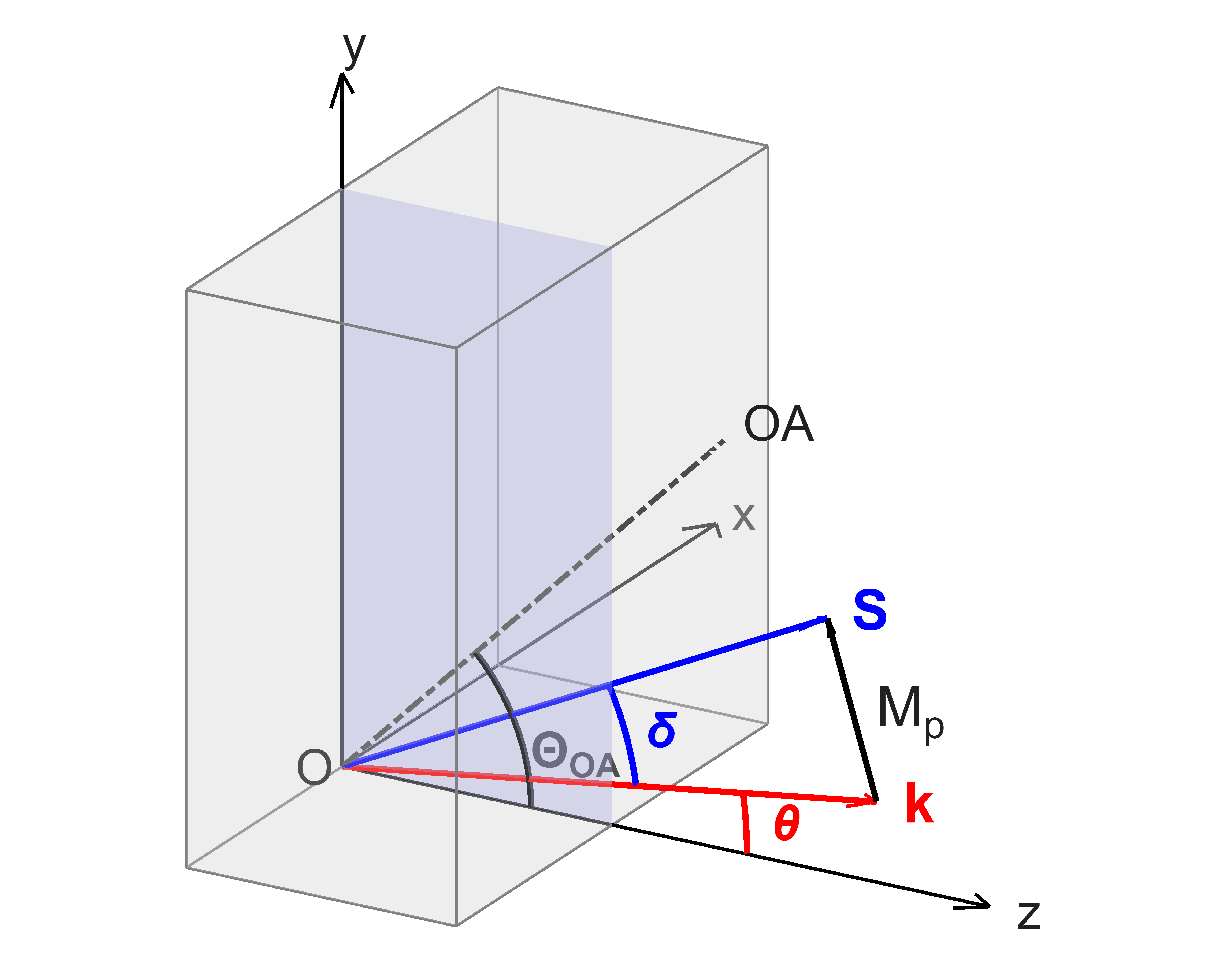}
\caption{Extraordinary beam propagation inside a negative uniaxial bi-refringent crystal. The pump wave vector and optic axis (OA) make an angle $\theta$ and $\bm{\Theta}_{OA}$, with respect to the propagation direction (z-axis), respectively. The Poynting vector $\bm{S}$ is tilted at an angle $\delta$ relative to $\bm{k}$. $\bm{M}_p$ denotes the spatial walk-off vector directed towards $\bm{S}$ from $\bm{k}$.} 
	\label{walk_off_geometry}
\end{figure}
For our scope, we consider a continuous single--frequency pump beam at $\Omega_p$, and a type--I process in a negative uniaxial birefringent crystal, that is, the signal and idler photons are both ordinary polarized while the pump beam is extraordinary polarized ($e\rightarrow o + o$). In this case only the pump wave vector depends on the transverse momentum, \textit{i.e.} the angle the wave vector forms with the optical axis~\cite{di2002entangled,rubin1996transverse}. For an extraordinary beam, the direction of energy flow (along the Poynting vector $\bm{S}$) is not parallel to the wave vector, $\bm{k}$. Hence, intensity profile of the beam walks sideways relative to its wavefront normal. Taking this into account, we introduce the spatial walk-off vector, $\bm{M_p}$, lying in the plane comprising of the
optic axis and the pump beam propagation direction, as shown in Fig. \ref{walk_off_geometry}. Therefore, in quasi--monochromatic ($|\nu_j|\ll\Omega_j$ with $\omega_j = \Omega_j + \nu_j$) and paraxial ($|\bm{q}_j|\ll |\bm{Q}_j|$ with $\bm{\kappa}_j  = \bm{Q}_j + \bm{q}_j$) approximations the longitudinal wave vectors read as
\begin{align}
	k_{p}  &\approx K_p + \bm{M}_p\cdot \bm{q}_p - \frac{{q}_p^2}{2K_p}
		\\
	k_{jz} &\approx K_{jz} + \frac{\nu_j}{u_{jz}} + \frac{\nu_j^2}{2a_{jz}} -\frac{\bm{q}_j\cdot(\bm{q}_j +2 \bm{Q}_j)}{2K_{jz}}-\frac{\bm{Q}_j^2}{2K_{jz}}\,
\end{align}

Here, $\bm{M}_p = \nabla k_p = N{\hat p}$, and $|N|= -d\ln[n_\text{e}(\Omega_p,\theta)]/d\theta$, evaluated at $\theta = \Theta_\text{OA}$ with $n_\text{e}$ being the extraordinary index of refraction. We note that ${\hat p}$ is the unit vector along the walk-off. We have also introduced the group velocity $u_{jz} = dk_{jz}/d\omega_j$, and the group acceleration $a_{jz} = d^2k_{jz}/d\omega^2$ evaluated at $\omega_j = \Omega_j$ and $\bm{\kappa}_j  = \bm{Q}_j$.
Finally, subject to the phase matching conditions, $\Omega_1 + \Omega_2 =\Omega_p$, $K_{1z} + K_{2z} = K_p$, and $ \bm{Q}_1+ \bm{Q}_2 =0$, the longitudinal phase mismatch up-to second order approximation becomes
\begin{align}
	\Delta \approx &- \frac{\nu_1}{u_{1z}}  - \frac{\nu_2}{u_{2z}}  -\frac{\nu_1^2}{2a_{1z}}  -\frac{\nu_2^2}{2a_{2z}} +
		\nonumber \\
	&+\bm{M}_p\cdot \bm{q}_p +\frac{\bm{q}_1\cdot \bm{Q}_1}{K_{1z}} +\frac{\bm{q}_2\cdot \bm{Q}_2}{K_{2z}} + \frac{\bm{Q}_1^2}{2K_{1z}}+\frac{\bm{Q}_2^2}{2K_{2z}}
		\nonumber\\
	&\qquad \qquad \qquad \qquad- \frac{1}{2}\left[ \frac{{q}_p^2}{K_p} - \frac{{q}_1^2}{K_{1z}} - \frac{{q}_2^2}{K_{2z}} \right]\,.
\end{align}

\subsubsection{Collinear, degenerate type--I phase matching}
Considering a collinear, degenerate type--I SPDC, that is $\bm{Q}_1= \bm{Q}_2= 0$, $\Omega_1 = \Omega_2 = \Omega_p/2$, $K_1 = K_2 =K_p/2$, and $\bm{q}_j = \bm{\kappa}_j$, the longitudinal phase mismatch, neglecting the second order frequency contribution, reads as
\begin{align}\label{eq:Delta}
    \Delta\approx
		 \bm{M}_p\cdot (\bm{\kappa}_1+\bm{\kappa}_2)
	+\frac{|\bm{\kappa}_1-\bm{\kappa}_2|^2}{2{K}_p}\,,
\end{align}
and the spectral function is provided by
\begin{align}\label{Phi}
	\phi(\bm{k}_{1},\bm{k}_{2}) \equiv  \widetilde\Phi(\bm{\kappa}_{1}&,\bm{\kappa}_{2}) =
		\widetilde{E_p}(\bm{\kappa}_{1}+\bm{\kappa}_{2})\times
			 \\
		&\times\int_{-L}^0 dz\,
		\e^{\ii \bm{M}_p\cdot (\bm{\kappa}_1+\bm{\kappa}_2)z}
		\e^{\ii\frac{|\bm{\kappa}_1-\bm{\kappa}_2|^2}{2{K}_p}z}
		\,.\nonumber
\end{align}
We note that the spectral function is not factorizable in the $ \bm{\kappa}_1,  \bm{\kappa}_2$ variables, neither, in general, in the transformed coordinates, $\bm{\kappa}_{+}=\bm{\kappa}_{1}+\bm{\kappa}_{2}$ and $\bm{\kappa}_{-}=\bm{\kappa}_{1}-\bm{\kappa}_{2}$.

\section{Field Evolution and detection}
The evolution of the quantum state of the down-converted field at the output face of the crystal, through a linear optical system, as described in Fig.~\ref{schematic}, can be carried out in Heisenberg picture where the operators need to be expressed in terms of their value at initial position. The electric field at the detectors plane can be written in terms of the electric field at the output surface of the crystal~\cite{rubin1994theory} as,
\begin{equation}\label{efield}
  	{E}_{j}^+ (x_j)=\sum_{\bm{k}_{j}}a_{\bm{k}_{j}}
			G_{j}(\bm{k}_j,x_j)\,,
\end{equation}
where $G_{j}(\bm{k}_j,x_j)$ is the optical system function, which describes the propagation of the output beam from the crystal plane to the detectors $1$ and $2$. Here, $x_j = (\bm{\rho}_{j},z_{j}, T_j)$, and  $T_{j}=t_{j}-z_{j}/c$, $t_{j}$, being the time when the $j-$th  detector reveals a photon.

The correlation function of the signal beam can be written as
\begin{align}\label{eq:G11}
   G^{(1)}_{11}(x_1,x_1^\prime)
   		&= \langle \psi |{E}_{1}^{(-)}{E}_{1}^{(+)}|\psi \rangle
   				\nonumber \\
		&= \sum_{\bm{k}_{2}} \big|\langle 0| a_{\bm{k}_{2}}{E}_{1}^{(+)}|\psi \rangle\big|^{2}
   			 	\nonumber\\
		&= \sum_{\bm{k}_{2}}\sum_{\bm{k}_{2}^\prime}\,\delta(\bm{k}_2-\bm{k}_2^\prime)\times
				\nonumber \\
		&\quad \qquad\times
				\sum_{\bm{k}_{1}}  G^\ast_1(\bm{k}_1,x_1) F^\ast(\bm{k}_{1},\bm{k}_{2})
				\nonumber\\
		&\quad \qquad\times \sum_{\bm{k}_{1}^\prime} G_1(\bm{k}_1^\prime,x_1^\prime) F(\bm{k}_{1}^\prime,\bm{k}_{2}^\prime)
				\,,
\end{align}
having a similar expression for the idler beam. The probability of detecting a single photon  in $x_1$  is $G_{11}^{(1)}(x_1,x_1)$. The probability to detect the photons pair in $x_1$ and $x_2$ is given by the correlation function
\begin{align}\label{eq:G12}
   G^{(2)}_{12}(x_1,x_2) = \langle &\psi |{E}_{1}^{(-)}{E}_{2}^{(-)}{E}_{2}^{(+)}{E}_{1}^{(+)}|\psi \rangle
   				=\big|A_{12}(x_1,x_2)\big|^{2}
				\nonumber \\
		&= \sum_{\bm{k}_{2}}\sum_{\bm{k}_{2}^\prime}G^\ast_2(\bm{k}_2,x_2) G_2(\bm{k}_2^\prime,x_2)\times
				\nonumber \\
		&\quad \qquad\times	\sum_{\bm{k}_{1}}  G^\ast_1(\bm{k}_1,x_1)
					F^\ast(\bm{k}_{1},\bm{k}_{2})
				\nonumber\\
		&\quad \qquad\times \sum_{\bm{k}_{1}^\prime} G_1(\bm{k}_1^\prime,x_1)
					F(\bm{k}_{1}^\prime,\bm{k}_{2}^\prime)\,,
\end{align}
where the biphoton amplitude is defined as
\begin{align}\label{eq:A12}
	A_{12}(x_1,x_2) &= \langle 0| {E}_{2}^{(+)}{E}_{1}^{(+)}|\psi \rangle
				   \\
				  & =  \sum_{\bm{k}_{1}} \sum_{\bm{k}_{2}}
				  	G_2(\bm{k}_2,x_2) G_1(\bm{k}_1,x_1)
					F(\bm{k}_{1},\bm{k}_{2})\,.
				\nonumber
\end{align}

\subsection{Free propagation}
In case of absence of any optical elements, the beam propagates freely and, in paraxial approximation, the optical system function  in Eq.~\eqref{efield} can be written as $G_{j}(\bm{k}_j,x_j)=g_{j}(\bm{\kappa}_{j},\omega_{j},\bm{\rho}_{j},z_{j})\,\exp{({-\ii\omega_{j}T_{j}})}$, where
\begin{align}
 	g_{j}(\bm{\kappa}_{j},\omega_{j},\bm{\rho}_{j},z_{j})= \e^{\ii \frac{\omega_j}{c}z_j}
			\psi\left(\bm{\kappa}_{j},\frac{-cz_j}{\omega_{j}}\right)
			\e^{\ii\kappa_j\cdot \bm{\rho}_{j}}
			\,,
\end{align}
with $z_j$ being the distance between the output crystal face and the $j-$th detection plane, and
\begin{align}
	h_{\omega}(\bm{\rho},d) &= \frac{-\ii\omega}{2\pi cd}\,\e^{\ii\frac{\omega}{c}d}\psi\left(\bm{\rho},\frac{\omega}{cd}\right)\,,
		\\
		\psi\left(\bm{\rho},\frac{\omega}{cd}\right) &= \e^{\ii\frac{\omega}{2cd}\bm{\rho}^{2}}\,.
\end{align}
Here, $h_{\omega}(\bm{\rho},d)$ is the impulse response function of a point source at a plane $\bm{\rho}$ after a propagation of distance $d$.

\subsection{Optical System Functions}
We now assume that a lens is placed  in a plane ($\bm{\rho}_{s}$) at a distance $d_{1}$ from the crystal plane ($\bm{\rho}_{x}$), as in Fig. \ref{schematic}. The detectors are $d_{1}^{\prime}$ further away from the lenses. The lens can be represented by an aperture function $L(\bm{\rho}_{s})$~\cite{goodman2005introduction}
\begin{align}
 	L(\bm{\rho}_{s}) &=e^{-i\frac{\omega}{2cf}\bm{\rho}_{s}^{2}}\,.
\end{align}
The corresponding optical system function, in paraxial approximation, becomes~\cite{rubin1996transverse,goodman2005introduction}
\begin{align}
	g_{j}(\bm{\kappa}_{j},\omega_{j},\bm{\rho}_{j},z_{j})
		=\int d^{2}\rho_{s}
		h_{\omega_{j}}(\bm{\rho}_{j}-\bm{\rho}_{s},d_{1}^{\prime})L(\bm{\rho}_{s})\times
			\nonumber \\		
		\times\int d^{2}\rho_{x}
		h_{\omega_{j}}(\bm{\rho}_{s}-\bm{\rho}_{x},d_{1})\,\e^{\ii\bm{\kappa}_{j}.\bm{\rho}_{x}}.
\end{align}
This is further simplified to 
\begin{align}
 	g_{j}(\bm{\kappa}_{j},\omega_{j},\bm{\rho}_{j},z_{j}) &=
		\frac{-\ii\omega_{j}}{2\pi cd_{1}^{\prime}} \,\e^{\ii\frac{\omega_{j}}{c}z_{1}}
			\psi\left(\bm{\kappa}_{j},-\frac{cd_{1}}{\omega_{j}}\right)\times
			  \\
		&\times \psi\left(\bm{\rho}_{j},\frac{\omega_{j}}{cd_{1}^{\prime}}\right)
		S_{1}\left(\bm{\kappa}_{j}-\frac{\omega_{j}}{cd_{1}^{\prime}}\bm{\rho}_{j},\frac{\omega_{j}}{cd_{1}^{\prime}}\right),
		\nonumber
\end{align}
where,
\begin{align}\label{eq22}
	S_{1}\left(\bm{\kappa},\frac{\omega}{cd}\right)
		&= \int d^{2}\rho_{s}\psi\left(\bm{\rho}_{s},\frac{\omega}{cd}\right)
				L(\bm{\rho}_{s})\,\e^{\ii\bm{\kappa} .\bm{\rho}_{s}}
				\nonumber \\
  		&= \int d^{2}\rho_{s}\psi\left[\bm{\rho}_{s},\frac{\omega}{c}\left(\frac{1}{d}-\frac{1}{f}\right)\right]\,
				\e^{\ii\bm{\kappa} .\bm{\rho}_{s}}.
\end{align}
Then, the explicit form of optical system function for specific cases needs to be evaluated to calculate the biphoton amplitude and we consider the following cases.

\subsubsection{Far Field}
In case both the crystal and the detector are kept at the focal plane of the lens, such that $d_{1}=d_{1}^{\prime}=f$, the corresponding optical system function for $j-$th detector is given by
\begin{align}
 	g_{j}(\bm{\kappa}_{j},\Omega_{j},\bm{\rho}_{j},2f)= \frac{-i\Omega_{j} \e^{\ii\frac{\Omega_{j}}{c}2f}}{2\pi cf}\,
			\delta\left(\bm{\kappa}_{j}-\frac{\Omega_{j}}{cf}\bm{\rho}_{j}\right).
\end{align}
Here, $\omega$ is replaced by $\Omega$ since $\omega\approx\Omega$. The single beam correlation function provided by Eq.~\eqref{eq:G11} can be cast as
\begin{align}\label{eq:G11_Far}
   G^{(1)}_{11}(\bm{\rho}_1,\bm{\rho}_1^\prime) \propto
		\iint d\bm{\kappa}_2\,
			&\widetilde\Phi^\ast\left(\frac{\Omega_1}{cf}\bm{\rho}_1,\bm{\kappa}_{2}\right)\times
				\nonumber \\
			\times&\widetilde\Phi\left(\frac{\Omega_1}{cf}\bm{\rho}_1^\prime,\bm{\kappa}_{2}\right)
				\,,
\end{align}
and the single counts are obtained by imposing $\bm{\rho}_1 = \bm{\rho}_1^\prime$, to yield $G^{(1)}_{11}(\bm{\rho}_1,\bm{\rho}_1)$ and the coincidence count reduces to
\begin{align}\label{eq:G12_Far}
   G^{(2)}_{12}(\bm{\rho}_1,\bm{\rho}_2)  \propto
   	\bigg|
		\widetilde\Phi\left(\frac{\Omega_1}{cf}\bm{\rho}_1,\frac{\Omega_2}{cf}\bm{\rho}_2\right)
	\bigg|^2.
\end{align}

\subsubsection{Near Field}
In this scenario, the lens in Fig. \ref{schematic} is placed in such a way that both the crystal and the detector are at $2f$ distance away from it: $d_{1}=d_{1}^{\prime}=2f$. As a result, the optical system function for $j$--th detector turns out to be
\begin{equation}
	g_{j}=-\e^{\ii\frac{\Omega_{j}}{c}4f}
			\psi \left(\bm{\rho}_{j},\frac{\Omega_{j}}{cf}\right)
			\,\e^{\ii\bm{\kappa}_{j}\cdot(-\bm{\rho}_{j})}\,.
\end{equation}
The single beam correlation function provided by Eq. ~\eqref{eq:G11} can be written as
\begin{align}\label{eq:G11_Near}
   G^{(1)}_{11}(\bm{\rho}_1,\bm{\rho}_1^\prime) & \propto
	\e^{\ii\frac{\Omega_1}{c2f}(\rho_1^2 - \rho_1^{\prime\,2})}
	\iint d\bm{\kappa}_2\times
			\nonumber \\
	&\times\iint d\bm{\kappa}_1^\prime 	\e^{-\ii\bm{\kappa}_1^\prime\cdot(-\bm{\rho}_1^\prime)}\,
							\widetilde\Phi^\ast(\bm{\kappa}_{1}^\prime,\bm{\kappa}_{2})
			\nonumber \\
	&\times\iint d\bm{\kappa}_1	 	\e^{\ii\bm{\kappa}_1\cdot(-\bm{\rho}_1})\,
							\widetilde\Phi(\bm{\kappa}_{1},\bm{\kappa}_{2})
				\,,
\end{align}
and the single counts are $G^{(1)}_{11}(\bm{\rho}_1,\bm{\rho}_1)$. The coincidence count can be calculated as
\begin{align}\label{eq:G12_Near}
   G^{(2)}_{12}((\bm{\rho}_1,\bm{\rho}_2)  \propto
   	\bigg|
		\iint d\bm{\kappa}_1 \iint d\bm{\kappa}_2 &\,
				\e^{\ii\bm{\kappa}_1\cdot(-\bm{\rho}_1) + \ii\bm{\kappa}_2\cdot(-\bm{\rho}_2})\times
		\nonumber \\
				&\times
				\widetilde\Phi(\bm{\kappa}_{1},\bm{\kappa}_{2})
	\bigg|^2.
\end{align}
To obtain the explicit form of $\widetilde\Phi(\bm{\kappa}_{1},\bm{\kappa}_{2})$, the field envelop has to be used as the finite energy Airy beam, which will be described in the following section.%
\section{Airy pump beam}
\begin{figure} [h!]
	\centering
	\includegraphics[angle=0, width=.45\textwidth]{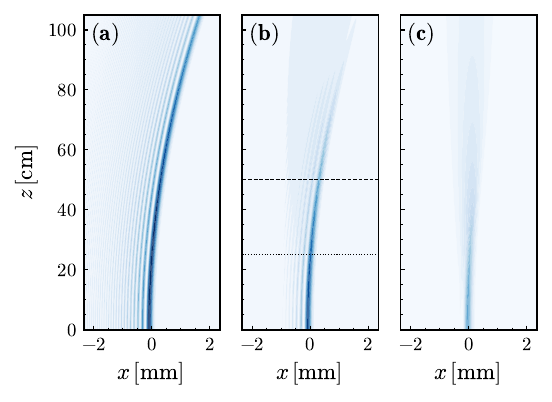}
	\caption{1D Airy pump beam propagation for  three values of the truncation parameter $w = (0.02,\,0.1,\,0.5)$, panel (a), (b) and (c), respectively.
		Panel (a) shows a diffracting-free propagation as anticipated in Ref.~\cite{berry1979nonspreading},  while panel (c) shows prominent diffraction.
		Other parameters for the beam propagation are $l = \SI{100}{\micro\meter}$, $\lambda = \SI{0.5}{\micro\meter}$~\cite{Siviloglou:07}.
		In panel (b) dotted and dashed lines mark the propagation distances at which the 2D profiles  in Fig.~\ref{fig:pump_beam_2D}  are evaluated.}
	\label{fig:pump_beam_1D}
\end{figure}

\begin{figure} [h!]
	\centering
	\includegraphics[angle=0, width=.48\textwidth]{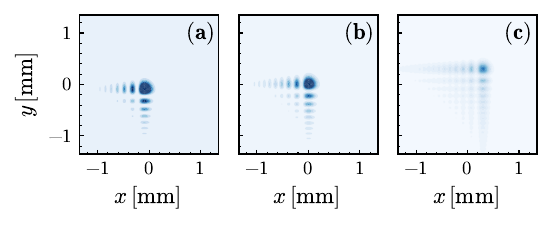}
	\caption{2D Airy pump beam profile for three values of the propagation distance $z = (0,\,25,\,50)\,\si{\centi\meter}$, panel (a), (b) and (c), respectively. They correspond to the propagation distances marked by dotted and dashed lines in panel (b) of Fig.~\ref{fig:pump_beam_1D}.}
	\label{fig:pump_beam_2D}
\end{figure}

For realizing a 1D finite energy Airy pump beam, an aperture function is used to truncate the infinite tail of Airy function. The truncated Airy beam takes the form
\begin{align}\label{eq:AiTr}
	\AiTr({x}/{l}) = \Ai({x}/{l})\,\e^{w{x}/{l}}\,,
\end{align}
where $\Ai(\xi)$ is the Airy function:
\begin{align}\label{eq:AiTrXi}
	\Ai(\xi) = \int \frac{dq}{2\pi}\,\e^{\ii q\xi}\, \e^{\ii q^3/3}\,.
\end{align}
In Eq.~\ref{eq:AiTr}, $w$ ($<1$) is the truncation parameter, $x$ is the transverse position coordinate, and $l$ is the characteristic length scale of the transverse field profile, for which the dimensionless variables are $\xi = x/l$ and $q = \kappa\,l$. We set  $l =\SI{100}{\micro\meter}$. The Fourier transform of the truncated profile takes the form
\begin{align}\label{eq:FTAiTr}
	\FTAiTr(q) = \e^{\ii (q + \ii w)^3/3}\,,
\end{align}
which shows a gaussian power spectrum and can be produced after the Fourier transformation of a Gaussian pump beam modulated by a cubic phase mask. The free propagation of a beam that has a truncated Airy amplitude at $z=0$, is given by~\cite{Siviloglou:07,siviloglou2007observation}
\begin{align}\label{eq:AiTrZ}
	\AiTr(\xi,\zeta) = \Ai\big[&\xi - (\zeta/2)^2 + \ii w\zeta\big] \times
			\nonumber \\
			&\times \e^{w\xi - w\zeta^2/2 - \ii\zeta^3/12 + \ii w^2\zeta/2 + \ii \xi \zeta/2},
\end{align}
where $\zeta =  z / (K_pl^2)$ is a normalized propagation distance.
In Fig.~\ref{fig:pump_beam_1D}, the propagation of an Airy pump beam is plotted for the values of the truncation parameter $w = (0.02,\,0.1,\,0.5)$. For a 2D finite energy Airy pump beam the initial envelope is given by
\begin{align}
	E_p (\bm{\rho})
		= \Ai\big({x}/{l}\big) \times
		   \Ai\big({y}/{l}\big)\,.
\end{align}
The beam profiles for $w=0.1$ at three propagation distances $z = (0,25,50)\,\si{\centi\meter}$ are shown in Fig.~\ref{fig:pump_beam_2D}.

The transverse contribution to the biphoton state is also factorized as
\begin{equation}
	\widetilde{E_p}(\bm{\kappa}_p)
		= \FTAiTr(\kappa_xl) \times  \FTAiTr(\kappa_yl)
		\,,
\end{equation}
for which the 2D problem reduces to two 1D ones. From now on, we assume $\hat x$ and $\hat y$, the orthogonal (ordinary) and parallel (extraordinary) directions, respectively, with respect to the plane comprising of the pump beam propagation direction, $\hat z$ and the crystal optical axis. We also assume a truncation parameter equal to $w=0.1$.

\subsection{Transverse momentum correlation}
The spectral function, $F(\bm{k}_{1},\bm{k}_{2})$, represents the scattering amplitude in the output modes, $\bm{k}_{1},\bm{k}_{2}$, and it is related to the probability density per unit frequency and per unit solid angle:
\begin{align}\label{eq:probability_density}
	\frac{dP(\bm{k}_{1},\bm{k}_{2})}{d\omega_1d\omega_2d\Omega_1d\Omega_2} \propto
		\left|
			F(\bm{k}_{1},\bm{k}_{2})
		\right|^2\,,
\end{align}
which provides the transverse momentum correlations of the biphoton state. We consider a collinear, degenerate type--I SPDC, for which the longitudinal phase matching is provided by Eq.\eqref{eq:Delta}. We note that the pump walk--off vector has component only along the $\hat y$ direction, that is $\bm{M}_p = N{\hat y}$, and we assume $N = 0.2$. The spectral function is obtained from $\widetilde\Phi({\kappa}_{1\alpha},{\kappa}_{2\alpha})$:
 \begin{align}\label{eq:SpectralFunction}
	\widetilde\Phi({\kappa}_{1\alpha},{\kappa}_{2\alpha})
		\propto \widetilde{E_p}({\kappa}_{1\alpha}+{\kappa}_{2\alpha})\,
		\text{sinc}\left(\frac{\Delta_\alpha L}{2}\right)\,
		\e^{-\ii \frac{\Delta_\alpha L}{2}}\,,
\end{align}
with $\alpha=(x,y)$, and $\Delta_\alpha$, provided in Eq.~\eqref{eq:Delta}. It is separable in $x$ and $y$ components, as long as the paraxial approximation is valid. It is to be noted that the biphoton amplitude can't be factorized into transverse momentum coordinates, ${\kappa}_{1\alpha}$ and ${\kappa}_{2\alpha}$, and hence, the photons are entangled in the spatial frequency domain. One way to analyze their entanglement is to find out the transverse correlation by evaluating the probability density in Eq.\eqref{eq:probability_density}.

\subsubsection{Spectral Function in the Ordinary Direction}
\begin{figure}[t]
    	\centering
	\includegraphics[angle=0, width=.495\textwidth]{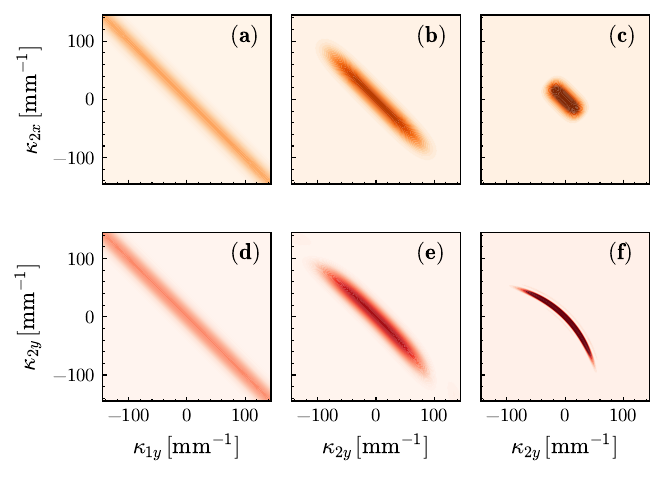}
	\caption{Probability density for the crystal lengths $L = (0.1,\,1,\,10)\,\si{\milli\meter}$ with truncation parameter $w=0.1$, for the ordinary and extraordinary components, panels (a--c) and (d--f), respectively.}
	\label{fig:correlations}
\end{figure}

In the $\hat x$-direction, the phase mismatch is given only in terms of the diffraction contribution:
\begin{align}\label{eq:Delta_ORD}
	\Delta_x =  \frac{({\kappa}_{1x}-{\kappa}_{2x})^2}{2{K}_p}\,,
\end{align}
and therefore, the biphoton in the transformed coordinate, ${\kappa}_{+}={\kappa}_{1x}+{\kappa}_{2x}$ and ${\kappa}_{-}={\kappa}_{1x}-{\kappa}_{2x}$, reads as $\widetilde\Phi({\kappa}_{+},{\kappa}_{-})\propto \text{sinc}\left(L\kappa_{-}^2/4K_p\right)\widetilde{E_p}({\kappa}_{+})$.
In Fig.~\ref{fig:correlations}(a--c), the probability densities of Eq.~\eqref{eq:probability_density} for three values of the crystal length$, L = (0.1,\,1,\,10)\,\si{\milli\meter}$ are shown. We note the anti--correlation of the transverse momentum of the two photons especially for short crystal length. However, increasing the crystal length, the momentum correlations localize due  to diffraction.

\subsubsection{Spectral Function in the Extraordinary Direction}
In the $\hat y$ direction, the phase mismatch also has a spatial walk--off contribution:
\begin{align}\label{eq:Delta_EXT}
	\Delta_y =N ({\kappa}_{1y}+{\kappa}_{2y}) + \frac{({\kappa}_{1y}-{\kappa}_{2y})^2}{2{K}_p}
\end{align}
In this case the biphoton does not factorize even in the transformed coordinate, ${\kappa}_{+}$ and ${\kappa}_{-}$. In Fig.~\ref{fig:correlations}(d--f), the probability densities of Eq.~\eqref{eq:probability_density} for three values of the crystal length, $ L = (0.1,\,1,\,10)\,\si{\milli\meter}$ are shown. We note again the anti--correlation of the transverse momentum for short crystal length. However, when the crystal length increases, the diffraction localizes the correlation, while the spatial walk--off effect bends the profile.

\begin{figure}[b]
    	\centering
	\includegraphics[angle=0, width=.495\textwidth]{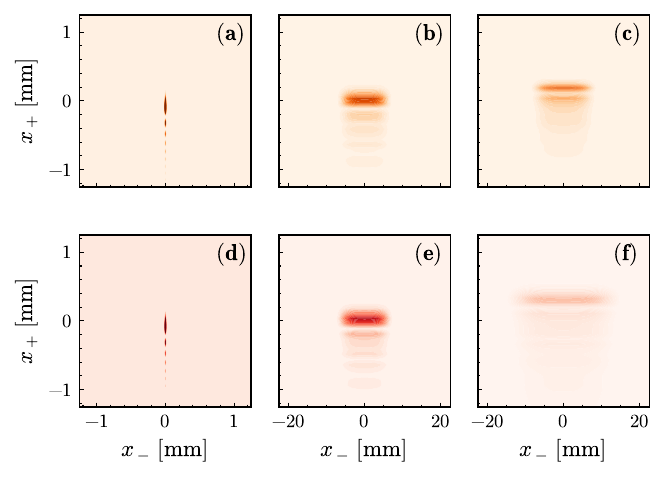}
	\caption{$|A_{12}|^2$ as a function of $x_+ = (x_1 + x_2)/2$ and $x_- = (x_1 - x_2)/2$ for three values of the propagation distance $z = (0,\,25,\,50)\,\si{\centi\meter}$, panel (a,d), (b,e) and (c,f), respectively. In panels (a,b,c) and (d,e,f) refer to the propagations of the ordinary and extraordinary components, respectively. The crystal length is $L= \SI{1}{\milli\meter}$.}
	\label{fig:Figure_Biphoton_vs_z}
\end{figure}

\subsection{Biphoton amplitude}
The biphoton amplitude defined in Eq.~\eqref{eq:A12} provides an insight on the photon pairs behaviour as a function of the propagation distance.
In case of free propagation, for which $G_j(\boldsymbol{\kappa}_j, z) \propto e^{-iz\left|\boldsymbol{\kappa}_j\right|^2/(2k_j)}$, the biphoton amplitude in terms of the normalized coordinates $\xi_j = (x_j/l, y_j/l)$ and propagation distance $\zeta = 2z/(K_pl^2)$ becomes
\begin{align}
	A_{12}(\xi_1,\xi_2; \zeta) &\propto
		 \int_{-1}^0 ds \,
		\int dq \,
		\e^{\ii q\xi_-}\,
		\e^{-\ii\frac{\zeta - sD}{2}q^2}\,		
		\times
			\nonumber \\
		&\hspace{10mm}\times
		\int dQ \,
		\e^{\ii Q(\xi_+ + s D_e)}\,
		\e^{-\ii \frac{ \zeta}{2}Q^2}\,
		\FTAiTr\big(Q\big)
			\nonumber \\
		&= \int_{-1}^{0} ds\, \sqrt{\frac{2\pi}{\zeta - sD}}\,\e^{\ii \frac{\xi_-^2}{2(\zeta - sD)}}\,\times
			\nonumber\\
		&\hspace{10mm}\times\,\AiTr\big[\xi_+ + sD_e,\zeta\big],
\end{align}
with  $Q = (\kappa_1 + \kappa_2)l$,  $q = (\kappa_1 - \kappa_2)l$, $\xi_+ = (\xi_1+\xi_2)/2l$, $\xi_-  = (\xi_1-\xi_2)/2l$, $D=L/(K_pl^2)$, and $D_e = NL/l$ for extraordinary coordinates only, otherwise becomes zero. The integration on the normalized variable $s$ over the range $[-1,0]$ reminds the integration over the crystal length.
In Fig.~\ref{fig:Figure_Biphoton_vs_z}, the modulus square of the biphoton amplitude  is reported for three values of the propagation distance, $z = (0,\,25,\,50)\,\si{\centi\meter}$, and for a crystal length equal to $L= \SI{1}{\milli\meter}$ and $w=0.1$.
We note that in the $x_+$ direction, the profile of the biphoton amplitude reshapes the truncated Airy pump beam profile, while it diffracts along $x_-$. The far field and near field limits will be obtained in the $f-f$ and $2f-2f$ configurations, corresponding to $d_1=d'_1=f$ and $d_1=d'_1=2f$, respectively.

\subsection{Far field}
We focus on the propagation of the state post SPDC which is generated by an 2D Airy pump beam through an $f-f$ optical system, which means that both the crystal and the detectors are kept at the focal plane of the lens such that $d_{1}=d_{1}^{\prime}=f$.
\subsubsection{Single counts}
\begin{figure}[b]
    	\centering
	\includegraphics[angle=0, width=.44\textwidth]{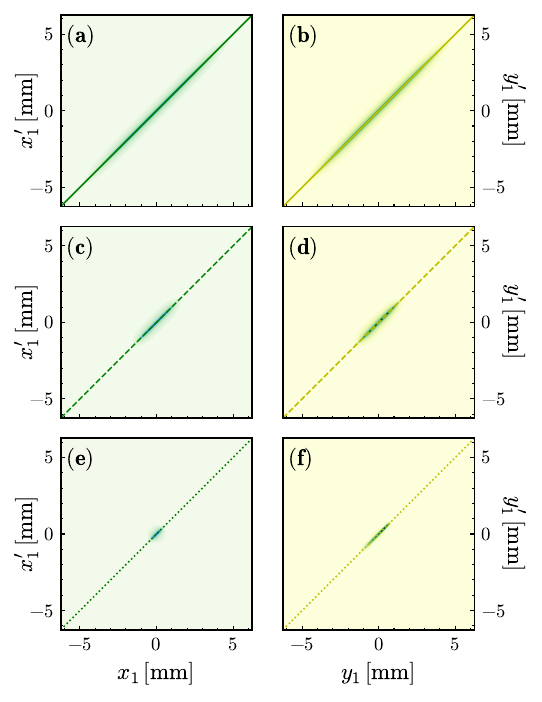}
	\caption{$|G_{11}^{(1)}|$ for far field as a function of the detector position for the ordinary component and for three values of the crystal length, $ L = (0.1,\,1,\,10)\,\si{\milli\meter}$, panel (a), (c) and (e), respectively. For the extraordinary component,  it is reported in panel (b), (d) and (f), respectively. The focal length is $f = \SI{100}{\milli\meter}$ and $w=0.1$.
	Solid, dashed and dotted lines mark the detection probabilities at $x_1 = x_1^\prime$ and $y_1 = y_1^\prime$,  which are  reported in Fig.~\ref{fig:FarField_G11_vs_G12}.}
	\label{fig:FarField_G11}
\end{figure}
Single beam correlation function in Eq.~\eqref{eq:G11_Far} for the ordinary direction, becomes
\begin{align}\label{eq:G11_Far_1D}
   G^{(1)}_{11}(x_1,x_1^\prime) & \propto
		\int d{\kappa}\,
			\widetilde\Phi^\ast\left(\frac{K_p}{2f}x_1,{\kappa}\right)
			\widetilde\Phi\left(\frac{K_p}{2f}x_1^\prime,{\kappa}\right)
				\,.
\end{align}
The analogous expression holds for the extraordinary one. In Fig.~\ref{fig:FarField_G11}, $|G_{11}^{(1)}|$ is reported for the ordinary and extraordinary directions, and for three values of the crystal length, $L = (0.1,\,1,\,10)\,\si{\milli\meter}$. The focal length is $f = \SI{100}{\milli\meter}$ and $w=0.1$. The probabilities to detect a single photon is obtained by setting $x_1 = x_1^\prime$:
\begin{align}\label{eq:G11_Far_1D_singles}
   G^{(1)}_{11}(x_1,x_1) & \propto
		\int d{\kappa}\,
			\left | \widetilde\Phi\left(\frac{K_p}{2f}x_1,{\kappa}\right)\right|^2
				\,,
\end{align}
that is, the diagonals in Fig.~\ref{fig:FarField_G11}, which are shown in Fig~\ref{fig:FarField_G11_vs_G12}.
We note that increasing the crystal length, the detection region decreases.

\begin{figure}[b]
    	\centering
	\includegraphics[angle=0, width=.44\textwidth]{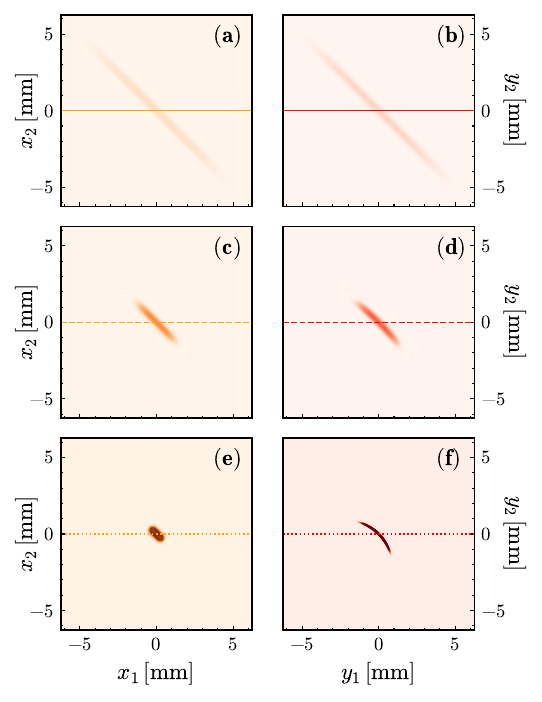}
	\caption{Normalized coincidences $G^{(2)}$ for far field and for the ordinary component, as a function of the detector position and for three values of the crystal length $ L = (0.1,\,1,\,10)\,\si{\milli\meter}$, panel (a), (c) and (e), respectively. For the extraordinary component, it is shown in panel (b), (d) and (f), respectively. The focal length is $f = \SI{100}{\milli\meter}$ and $w=0.1$.
	Solid, dashed and dotted  lines mark the conditional detection probabilities at $x_2 = y_2 = 0$, which are reported in Fig.~\ref{fig:FarField_G11_vs_G12}.}
	\label{fig:FarField_G12}
\end{figure}

\subsubsection{Coincidence counts}
Coincidence counts in Eq.~\eqref{eq:G12_Far}  becomes
\begin{align}\label{eq:G12_Far_1D}
   G^{(2)}_{12}(x_1,x_2) & \propto
			\left | \widetilde\Phi\left(\frac{K_p}{2f}x_1,\frac{K_p}{2f}x_2\right)\right|^2
				\,,
\end{align}
and are reported in Fig.~\ref{fig:FarField_G12} for the ordinary and extraordinary directions with three values of the crystal length, $L = (0.1,\,1,\,10)\,\si{\milli\meter}$ for the focal length $f = \SI{100}{\milli\meter}$ and $w=0.1$.
As expected, the shape of the coincidence counts resemble the probability density of the biphoton in the wave vector  space. We note that the shape for the extraordinary components is affected by the spatial walk--off of the pump beam.

\begin{figure}[h]
    	\centering
	\includegraphics[angle=0, width=.45\textwidth]{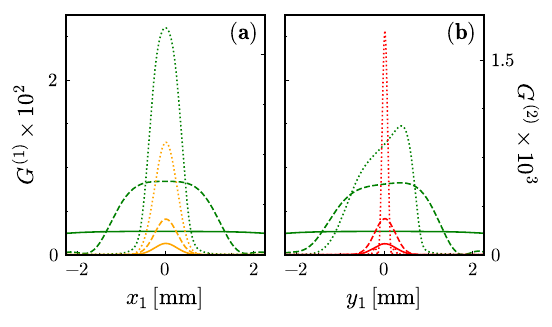}
	\caption{Normalized conditional coincidences, $G^{(2)}$, and single probabilities, $G^{(1)}$, for far field  as a function of the detector position for three values of the crystal length, $L = (0.1,\,1,\,10)\,\si{\milli\meter}$, represented by the solid, dashed and dotted lines, respectively. Panel (a) and (b) refer to the ordinary and extraordinary components, respectively. The focal length is $f = \SI{100}{\milli\meter}$ and $w=0.1$. In panel (a) and (b) green color corresponds to $G^1$ and orange, red colors correspond to $G^2$.}
	\label{fig:FarField_G11_vs_G12}
\end{figure}

In Fig.~\ref{fig:FarField_G11_vs_G12}, we compare the conditional coincidence probabilities
\begin{align}\label{eq:G2}
	G^{(2)} = \frac{G^{(2)}_{12}(x_1,0)}{\iint dx_1dx_2\,G^{(2)}_{12}(x_1,x_2)}\,,
\end{align}
and the single counts probabilities
\begin{align}\label{eq:G1}
	G^{(1)} = \frac{G^{(1)}_{11}(x_1,x_1)}{\iint dx_1\,G^{(1)}_{11}(x_1,x_1)}\,,
\end{align}
for the crystal lengths, $L=(0.1,\,1,\,10)\,\si{\milli\meter}$, and the focal length, $f =\SI{100}{\milli\meter}$.
It is observed that, upon increasing the crystal length, the phase--matching modifies the probabilities and for the extraordinary direction, it introduces an asymmetry in the single counts. The results for shorter crystal length are consistent with the spectral function expressed only by Fourier transform of the truncated Airy pump beam,
\begin{align}\label{eq:G12_Far_1DThin}
   G^{(2)}_{12}(x_1,x_2) & \propto
			\left| \FTAiTr\left[\frac{K_p\,l(x_1 + x_2)}{2f}\right]\right|^2
				\,,
\end{align}
while $G^{(1)}_{11}(x_1,x_1)$ becomes almost constant.

\subsection{Near field}
We focus on the propagation of SPDC generated by an 2D Airy pump beam through a $2f-2f$ optical system, that is both the crystal and the detectors are kept at twice the focal length of the lens: $d_{1}=d_{1}^{\prime}=2f$.

\begin{figure}[b]
    	\centering
	\includegraphics[angle=0, width=.45\textwidth]{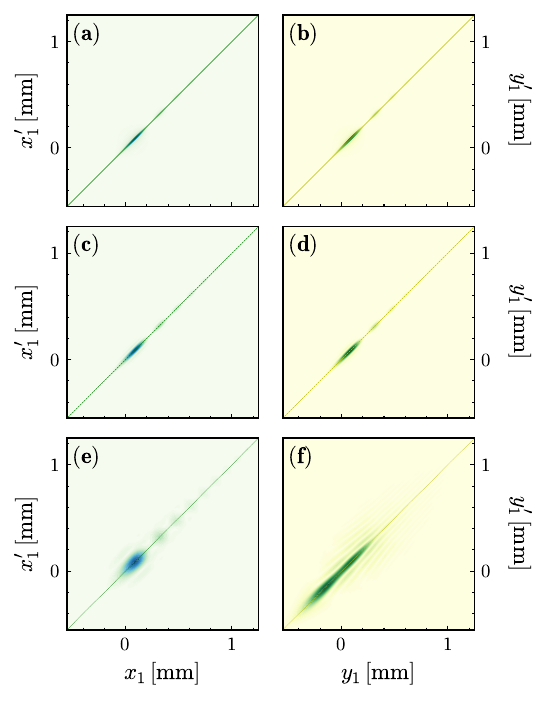}
	\caption{$|G_{11}^{(1)}|$ for near field and for the ordinary component as a function of the detector position for three values of the crystal length $ L = (0.1,\,1,\,10)\,\si{\milli\meter}$ in panel (a), (c) and (e), respectively and $w=0.1$. For the extraordinary component it is shown in panel (b), (d) and (f), respectively. The focal length is $f = \SI{100}{\milli\meter}$ and $w=0.1$.
	Solid, dashed and dotted lines mark the detection probabilities at $x_1 = x_1^\prime$ and $y_1 = y_1^\prime$, which are reported in Fig.~\ref{fig:NearField_G11_vs_G12}.}
	\label{fig:NearField_G11}
\end{figure}
\subsubsection{Single counts}
Single beam correlation functions in Eq.~\eqref{eq:G11_Near} can be written as
\begin{align}\label{eq:G11_Near_1D}
   G^{(1)}_{11}(x_1,x_1^\prime)  \propto
   		\e^{\ii \frac{K_p}{2f}(x_1^2 - x_1^{\prime\,2})}& \times
				 \\
			\times
			\int d{\kappa}& \int d{\kappa_1^\prime}\, \e^{-\ii \kappa_1^\prime (-x_1^\prime)}
				\widetilde\Phi^\ast\left(\kappa_1^\prime,{\kappa}\right)\,
				\nonumber \\
			\times
			&\int d{\kappa_1}\, \e^{\ii \kappa_1(-x_1)}
				\widetilde\Phi\left(\kappa_1,{\kappa}\right)\,,
				\nonumber
\end{align}
and are depicted in Fig.~\ref{fig:NearField_G11} for the same three values of the crystal length, $L = (0.1,\,1,\,10)\,\si{\milli\meter}$, the focal length $f = \SI{100}{\milli\meter}$ and $w=0.1$. The probabilities to detect a single photon become
\begin{align}\label{eq:G11_Near_1D_singles}
   G^{(1)}_{11}(x_1,x_1) & \propto
		\int d{\kappa}\,
			\left |
			\int d{\kappa_1} \e^{\ii \kappa_1(-x_1)}
				\widetilde\Phi\left(\kappa_1,{\kappa}\right)
			\right|^2\,,
\end{align}
as shown in the diagonals in Fig.~\ref{fig:NearField_G11} and they are explicitly shown in Fig~\ref{fig:NearField_G11_vs_G12}. It can be observed that, upon increasing the crystal length, the detection region spreads out, and the extraordinary component also gradually washes out.
\begin{figure}[b]
    	\centering
	\includegraphics[angle=0, width=.45\textwidth]{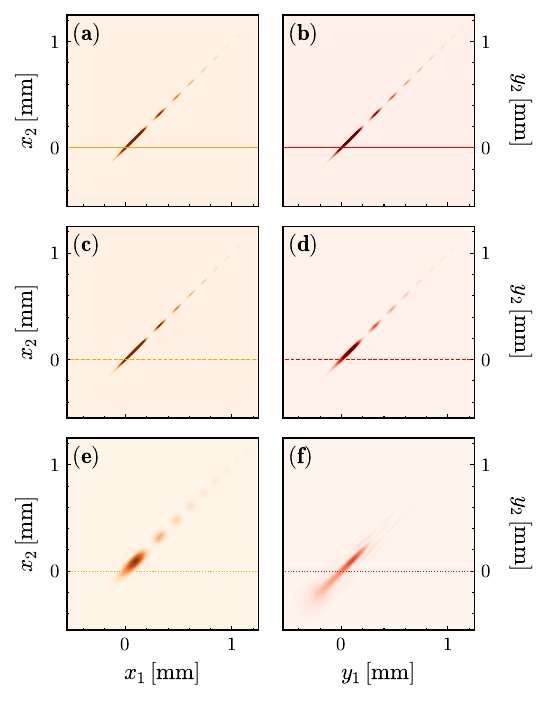}
	\caption{Near field normalized coincidences for the ordinary component as a function of the detector position for three values of the crystal length $ L = (0.1,\,1,\,10)\,\si{\milli\meter}$ in panel (a), (c) and (e), respectively. For the extraordinary component  it is reported in panel (b), (d) and (f), respectively. The focal length is $f = \SI{100}{\milli\meter}$ and $w=0.1$.
	Solid, dashed and dotted  lines mark the conditional detection probabilities at $x_2 = y_2 = 0$, which are reported in Fig.~\ref{fig:NearField_G11_vs_G12}.}
	\label{fig:NearField_G12}
\end{figure}
\subsubsection{Coincidence counts}
The coincidence counts in Eq.~\eqref{eq:G12_Far} takes the form
\begin{align}\label{eq:G12_Near_1D}
   G^{(2)}_{12}(x_1,x_2) & \propto
			\left |
			\iint d{\kappa_1}\,
			d{\kappa_2} \e^{-\ii \kappa_1x_1 -\ii \kappa_2x_2}
				\widetilde\Phi\left(\kappa_1,\kappa_2\right)
			\right|^2\,,
\end{align}
which are displayed in Fig.~\ref{fig:NearField_G12} for both the ordinary and extraordinary directions, and for three values of the crystal length, $L = (0.1,\,1,\,10)\,\si{\milli\meter}$, the focal length, $f = \SI{100}{\milli\meter}$ and $w=0.1$. As expected, the shape of the coincidence counts resemble the truncated Airy beam profile. For the extraordinary component, the spatial walk--off of the pump beam gradually washes out the Airy profile.

\begin{figure}[t]
    	\centering
	\includegraphics[angle=0, width=.44\textwidth]{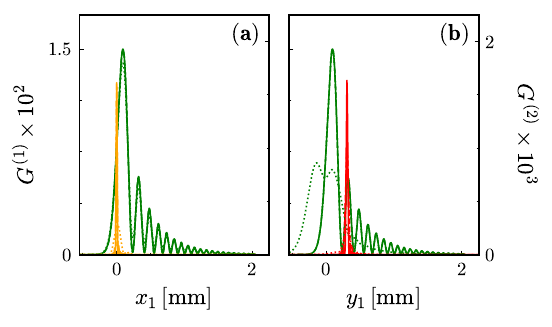}
	\caption{Normalized conditional coincidences, $G^{(2)}$, and singles probability, $G^{(1)}$, for near field as a function of the detector position for three values of the crystal length $L = (0.1,\,1,\,10)\,\si{\milli\meter}$, solid, dashed and dotted lines, respectively. Panel (a) and (b) refer to the ordinary and extraordinary components, respectively. The focal length is $f = \SI{100}{\milli\meter}$ and $w=0.1$. In panel (a) and (b), green color corresponds to $G^1$ and orange, red colors correspond to $G^2$.}
	\label{fig:NearField_G11_vs_G12}
\end{figure}

In Fig.~\ref{fig:NearField_G11_vs_G12}, we compare the conditional coincidence probabilities $G^{(2)}$ and the single counts probabilities $G^{(1)}$.
The results for shorter crystal length are consistent with the spectral function expressed only by the truncated Airy pump beam profile,
\begin{align}\label{eq:G12_Near_1DThin}
   G^{(2)}_{12}(x_1,x_2) & \propto
			{\rm Ai}^2\left[\frac{-2(x_1 + x_2)}{l}\right] \,\delta(x_1 - x_2)
				\,.
\end{align}
In this case, the single beam correlation function is given by
\begin{align}\label{eq:G11p_Near_1D}
   G^{(1)}_{11^\prime}(x_1,x_1^\prime) & \propto
			{\rm Ai}\left(-\frac{x_1}{l}\right)
			{\rm Ai}\left(-\frac{x_1^\prime}{l}\right) \,\delta(x_1 - x_1^\prime)
				\,,
\end{align}
and finally we obtain
\begin{align}\label{eq:G11_Near_1D1}
   G^{(1)}_{11}(x_1,x_1) & \propto
			{\rm Ai}^2\left(-\frac{x_1}{l}\right)
				\,.
\end{align}

\section{Spatial entanglement and Correlation}

The degree of spatial entanglement in type-I SPDC can be measured using Schmidt mode analysis of the two-photon coincidence distribution. The biphoton wavefunction which is proportional to $\widetilde\Phi(x_1,x_2)$ may be expressed in the Schmidt decomposition:
\begin{equation}
    \widetilde\Phi(x_1,x_2) = \sum_{n} \sqrt{\lambda_n} \, u_n(x_1)\, v_n(x_2),
\end{equation}
where $\{u_n(x_1)\}$ and $\{v_n(x_2)\}$ are orthonormal Schmidt modes for the signal and idler photons, respectively, and $\lambda_n$ is the corresponding Schmidt weight. The normalized coincidence distribution is given by
\begin{equation}
  P(x_1, x_2) \propto G^{(2)}_{12}(\bm{\rho}_1, \bm{\rho}_2),
\end{equation}
which serves as the input to the numerical singular value decomposition (SVD). The dimensionality of entanglement is captured by the effective Schmidt number,
\begin{equation}
    K = \frac{1}{\sum_n \lambda_n^2},
\end{equation}
while the mode entropy is written as
\begin{equation}
    S = -\sum_n \lambda_n \ln \lambda_n,
\end{equation}
with the single-photon purity
\begin{equation}
    \gamma = \sum_n \lambda_n^2.
\end{equation}
Physically, $K$ tells us how many independent spatial modes significantly contribute to the two-photon state. A small K means that the biphoton field is mostly made up of a few Schmidt modes (low-dimensional entanglement), while a large K means there exists highly multimode correlations. To comprehend the dimensionality and correlation strength of spatial entanglement, we will subsequently examine the Schmidt weights ($\lambda_n$) and cumulative mode weights ($\sum_n \lambda_n$) for both near- and far-field normalized coincidence distributions.

In our analysis, we use SVD directly on $\sqrt{P(x_1,x_2)}$ to perform the numerical Schmidt decomposition. The coincidence distribution is obtained from Eq. (\ref{eq:G12_Far}) and Eq. (\ref{eq:G12_Near}) by taking the experimentally relevant parameters: $w=0.05$, $f=100$ mm, $\lambda= \SI{0.5}{\micro \meter}$, $l=\SI{100}{\micro \meter}$ and $M_p=0.2 \hat{y}$. We ensure that the singular values ($\{s_n\}$) give Schmidt weights, $\lambda_n = s_n^2 / \sum_m s_m^2$. This method offers a good numerical estimate of $K$, $S$, and $\gamma$, for both the near field and the far field situations by considering the spectral function in the extra-oridinary direction.

\begin{figure*}[t]
    \centering
    \includegraphics[width=0.8\textwidth]{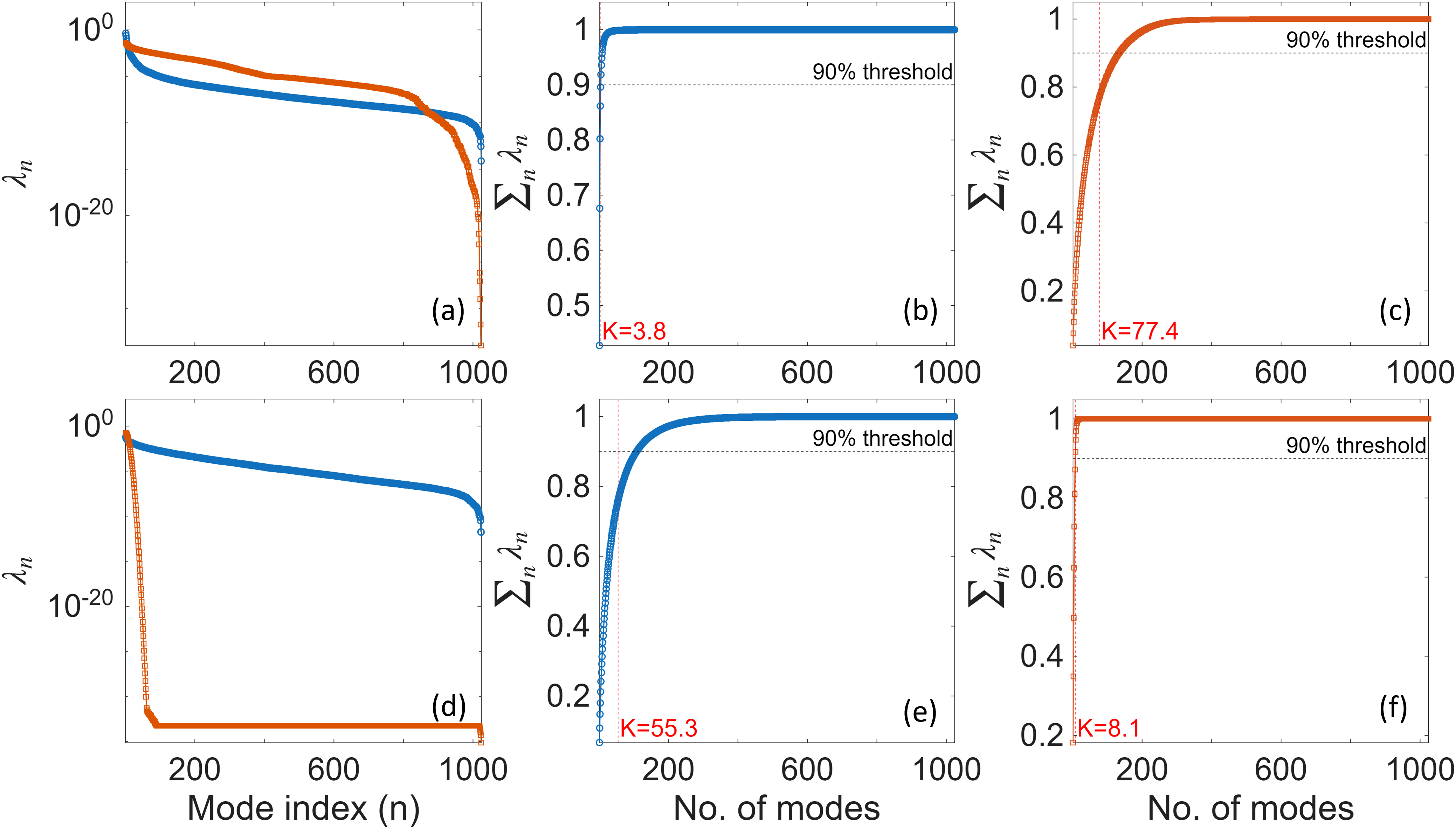}

    \caption{Schmidt weights ($\lambda_n$ in log scale) and cumulative mode weights ($\sum_n \lambda_n$) for long ($L=10$ mm) crystal ((a), (b) and (c)). For short crystal ($L=0.1$ mm), they are shown in (d), (e) and (f). Here, blue and red curves correspond to near- and far-field cases, respectively. The red-dotted vertical line indicates effective Schmidt number $K$ in each case.}
    \label{fig:schmidt_spectra}
    \end{figure*}

For the long crystal ($L = 10$ mm), the Schmidt spectrum in the near-field drops off very quickly, giving a small value of $K$ ($\approx 4$). This proves that the position correlations are strong and effectively low-dimensional. On the other hand, the decay is much slower in the far--field case and we get a large value of $K$ ($\approx 77$). This shows that the momentum distribution is highly multi-mode. The longer crystal makes sure that the phase matching is much stricter, which keeps the photon positions tightly locked in the near-field, but allows for a wide range of transverse momenta, creating a high-dimensional momentum entanglement. The corresponding entropies become $S_{\mathrm{NF}} \approx 1.8$ nats and $S_{\mathrm{FF}} \approx 4.9$ nats, which conforms to this trend. The purity, on the other hand, decreases from $\gamma_{\mathrm{NF}} \approx 0.27$ to $\gamma_{\mathrm{FF}} \approx 0.013$. This demonstrates how near- and far-field structures change from being almost single-mode to being highly multimode.

The observed trend is opposite for a short crystal ($L = 0.1$ mm for instance). In the near-field, the spectrum is wide and slowly fades away, resulting in a value of $K \approx 55$. This means that position locking is weaker and there are multi-mode spatial correlations. In the far-field, the decay is steeper, and only a few modes are needed ($K \approx 8$). This is in line with stronger momentum correlations when phase-matching is less strict. Physically, the short crystal looses the phase-matching constraint, which makes position space correlations weaker (multi-mode) and momentum space correlations sharper (few modes). The entropy and purity of the state also indicate a similar behavior, as we see the transition from $S_{\mathrm{NF}} \approx 4.6$ nats to $S_{\mathrm{FF}} \approx 2.2$ nats, accompanied by an increase in purity from $\gamma_{\mathrm{NF}} \approx 0.018$ to $\gamma_{\mathrm{FF}} \approx 0.12$.

Although the Schmidt number $K$ measures the effective dimensionality of entanglement, it is not a precise measure of the number of modes required to capture a given percentage of correlations. For longer crystal, only a few modes are needed in the near-field to capture $90\%$ of the correlations, whereas hundreds of modes are needed in the far-field. The opposite is true for shorter crystals, as it is shown in Fig. \ref{fig:schmidt_spectra}.

The Schmidt number depends on the relative widths of the pump envelope and the crystal's phase-matching function and these numerical results are consistent with analytical models based on Gaussian phase-matching approximations \cite{law2004analysis,miatto2012spatial}. Here, the effect of crystal length, which is embedded in the phase matching function, is emphasized since the Schmidt number changes dramatically by altering it.

\section{Conclusions}

In this work, we have considered an Airy beam propagation through a non-centrosymmetric birefringent crystal. After the down-conversion process, emergent photons are made to traverse through a convex lens before detection. For a type-I degenerate down-conversion, we derive the biphoton amplitude or coincidence count for both near- and far-field configurations. We see that the former case manifests the signature of finite-energy Airy beam which is a clear indication that the photon pairs remain correlated after the process. The coincidence count for far-field reflects a Gaussian spectrum which clearly maps the biphoton momentum distribution and shows its anti-correlation nature. We have also taken into account the Poynting vector walk-off and have shown its effect on the biphoton spectral functions and consequently on the joint probability densities. The probability density in the extraordinary direction bends its shape as a consequence of spatial walk-off innate to the extraordinary pump beam. The spectral function or the probability density also localizes with increasing crystal length due to diffraction. The single count probabilities suffer an asymmetry in the extraordinary direction because of the modified phase matching condition. This analysis helps us gain insight into the spectral properties of the entangled Airy biphoton, which may help us to generate a more robust source of entangled photons. In addition, by choosing an Airy beam as an input to SPDC and later recovering the original Airy pattern in the coincidence distribution may be useful for efficient quantum imaging \cite{monken1998transfer,rubin2008resolution,yue2025quantum}, due to its self-healing property. The analysis of the Schmidt number ($K$) further emphasizes the importance of experimental geometry by demonstrating that either long crystals in momentum space (far-field), or thin crystals in position space (near-field) can be used to engineer high-dimensional entanglement. Such control over the spatial Schmidt number is crucial for applications in quantum imaging, quantum communication, and continuous-variable quantum information processing.

\section*{Acknowledgement}

V. Sau thanks DST Inspire, Government of India, for financial support. P. Piergentili, D. Vitali and G. Di Giuseppe acknowledge financial support from NQSTI within PNRR MUR Project PE0000023-NQSTI.

\section*{Data Availability}
The data are available from the authors upon reasonable request.
\bibliography{reference_final}

\end{document}